\documentclass[reprint, prper]{revtex4-2}
\usepackage{comment}
\setcitestyle{numbers,square}
\usepackage{array}
\usepackage{amsmath}
\usepackage{units}
\usepackage{graphicx}
\usepackage{braket}
\usepackage{xfrac}
\usepackage{enumitem}
\usepackage[normalem]{ulem}
\usepackage{booktabs}
\usepackage{tabularx}
\usepackage{tikz}
\usepackage{multirow, bigstrut}
\usepackage{hhline}
\usepackage{color,soul}

\usepackage{times}

\usetikzlibrary{shapes.geometric, arrows, matrix, positioning}
\usetikzlibrary{fit}
\tikzset{%
highlight/.style={rectangle,rounded corners,draw,
fill opacity=0.5,thick,inner sep=0pt}
}

\usepackage{hyperref}
\hypersetup{
colorlinks=true,
linkcolor=cyan,
filecolor=magenta, 
urlcolor=blue,
}

\begin{document}

\title{Do we have a quantum computer?
Expert perspectives on current status and future prospects}

\author{Liam Doyle, Fargol Seifollahi and Chandralekha Singh}

\affiliation{
Department of Physics and Astronomy, University of Pittsburgh, Pittsburgh, PA, 15260 USA}


\begin{abstract}
The rapid growth of quantum information science and technology (QIST) in the 21st century has created both excitement and uncertainty about the field's trajectory. This qualitative study presents perspectives from leading quantum researchers, who are educators, on fundamental questions frequently posed by students, the public, and the media regarding QIST. Through in-depth interviews, we explored several issues related to QIST including the following  key areas: the current state of quantum computing in the noisy intermediate-scale quantum (NISQ) era and timelines for fault-tolerant quantum computers, including a possible timeline for quantum advantage on Shor's factoring algorithm, the feasibility of personal quantum computers in our pockets, and promising qubit architectures for future development. Our findings reveal diverse yet convergent perspectives on these critical issues. 
While experts agree that the current machines with physical qubits that are being built in the NISQ era should be called quantum computers, most estimated that it will take a decade to build a small fault-tolerant quantum computer, and  several decades to achieve scalable systems capable of running Shor's factoring algorithm with quantum advantage
and some stated that they would find it exciting if some law of physics prevents achieving those ultimate goals.
Regarding carrying a quantum computer in the pocket, 
experts 
viewed quantum computers as specialized tools that will remain in 
central locations such as data centers and can be accessed remotely for applications for which they are particularly effective compared to classical computers. 
Quantum researchers suggested that multiple platforms show promise, 
including neutral atoms, 
superconducting circuits, semiconducting qubits, and photonic systems, with no clear winner emerging. 
These insights provide valuable guidance for educators, policymakers, and the broader community in establishing realistic expectations for developments in this exciting field.
Our findings can provide valuable 
information for educators 
to clarify student doubts about these important yet confusing issues related to quantum technologies at a time we are celebrating the International Year of Quantum Science and Technology.

\end{abstract}

\maketitle

\section{Introduction}

The 21st century has started with a rapid growth in quantum information science and technology (QIST), an interdisciplinary field which promises disruptive transformations in our ability to compute, communicate and sense, taking advantage of quantum superposition and entanglement \cite{european,raymer2019,flagship,alexeev2021quantum,divincenzo,lloyd,daley2022,altmansimulation,logicalqubit,simulationfeynman,advantage,bennett1984,ekert1991quantum,kwiatqkd,RevModPhysqkd,qkdpanchina2017,entanglementphysics}. The groundbreaking theoretical work of Shor 
in the mid-1990s in both factoring (products of large prime numbers in polynomial time ~\cite{shor1994}, which makes the Rivest-Shamir-Adleman (RSA) protocol commonly used for encryption vulnerable) as well as quantum error-correction \cite{shorcode}, along with advances in our ability to control and manipulate microscopic systems, 
gave a major impetus to the field. In the last three decades, the field has evolved from purely theoretical concepts to physical implementations, with multiple technology companies and research institutions now operating quantum processors 
with different qubit architectures \cite{levymrs, trapped,semiconductorqubit,2020superconducting,circuitqed,neutralatom,kimble,awschalom,photon,photonic,nvcenter1}.

The current state of quantum computing is often characterized as the noisy intermediate-scale quantum (NISQ) era, a term coined by Preskill ~\cite{Preskill} to describe quantum processors
that can perform certain computational tasks beyond the reach of classical computers but remain susceptible to decoherence and errors. This intermediate stage between early proof-of-concept demonstrations and the goal of building fault-tolerant quantum computers, including the ultimate goal of scalable quantum computers has created both excitement \cite{fox2020cu, singhasfaw2021pt,meyer2022cu, asfaw2022ieee,muller2023prperworkforce, qtmerzeletal,bitzenbauer3,nvcenter2,jeremytpt,jeremyajp,kellyqkd,netoqkdoutreach,galvezqkd,Kohnleqkd,woottersqkd} and confusion among students 
and the broader public interested in this field.

The rapid progress in QIST has been accompanied by significant media attention, public interest, and substantial government and private sector investments. However, this attention has also led to misinformation  \cite{kashyap2025strategies} about the current capabilities of quantum computers, unrealistic timelines for practical applications, and confusion about which technological approaches (if any) show  
significant promise. Unlike the current classical computers built using transistors, which rely on a mature semiconductor technology as their backbone, at this early stage of the evolution of QIST, the proliferation of different qubit platforms--including superconducting circuits, neutral atoms, trapped ions, semiconducting qubits, and photonic systems--has further complicated students' as well as the public's understanding of the field's trajectory.

There has been extensive research on student understanding of quantum mechanics over the last 25 years, e.g., see Refs. \cite{singh2001,zollman2002,beckphoton,singh2008grad,singh2008interactive,zhu2012measure1,zhu2012measure2,zhu2012QMS,kohnle2013,singh2015review,marshman2015,marshman2019qmfps,emigh2015time,emigh2018,michelini2023research,chhabra2023undergraduate,marshman2017opejp,brown2016prper,marshman2017expect,sayerajp2025ltp}, and this work informs more recent education research focused on two state systems and QIST, e.g., see Refs. \cite{singh2007comp,marshman2016ejpphoton,Kohnle_2017,devore2020qkd,maries2020mzidouble,kiko,singh2022tpt,qtmerzel,qtbrang,qthellstern,qtgoorney,qtmeyercu,qtsun,hubloch,hucomputing,lopez2020encrypt,schalkers2024explaining,QISresource,galvez2014resource,meyer2023media,justicemathphysics}. 
Educating students \cite{rodriguez2020designing,donhauser2024empirical,Benlarmorajp2025,borish2025affordances,didics2015analysis,faletivc2025analogies,borish2025affordances,didics2015analysis,faletivc2025analogies,weissmanphysics,michelini2022,hennig2024new,hu2022time,hu2022uncertainty,hu2023changebasis,hu2023prper} and clarifying their doubts, as well as responding to common queries from the public and media regarding the status of this interdisciplinary QIST field is vital for ensuring that 
they have vetted information and a realistic timeline for the growth of this exciting field. Educators communicating about quantum technologies, both in formal and informal educational settings, face unique challenges in conveying both the revolutionary potential of quantum technologies \cite{goorney2024framework,bungum2022quantum,bondani} and the significant technical hurdles that remain in harnessing the envisioned quantum advantage. They must balance enthusiasm for the field's possibilities \cite{aaronson2013,nielsen2010quantum,mermin,raymerbook,wong,dancing,quantumoptics} with realistic assessments of current limitations and future timelines.

This paper focuses on the reflections of leading quantum researchers, who are also educators and who are passionate about teaching quantum, and about their perspectives on common questions that students, the public, and the media have about the field. By capturing expert perspectives on fundamental questions about quantum computing's current state, future prospects, and practical limitations, we aim to provide a valuable resource for educators, policymakers, and science communicators interested in QIST. In particular, the findings can be invaluable for other educators who are asked similar questions while teaching, giving public presentations, and talking to the media as well as for the development of curricula.

Our study addresses critical questions that frequently arise in formal educational settings as well as during informal public outreach activities: whether we truly have quantum computers in the NISQ era, what the realistic timelines are for small fault-tolerant quantum computers and scalable systems that will show quantum advantage on Shor's factoring algorithm,
and whether quantum computers will ever become personal devices. Additionally, we report quantum educators' perspectives on the qubit architectures they are most excited about and 
think would show great promise for the future development of quantum technologies. These questions reflect fundamental concerns about the field's current status, technological pathways, and its ultimate impact on society that should be communicated clearly to students in both formal and informal settings.

\section{Methodology}

Eighteen individuals, who are university faculty and conduct research in quantum-related fields with implications for QIST, were contacted via email or in person to request an interview. The selection process was based on the experts’ active involvement in QIST-related research and quantum education, along with their existing familiarity with one of the authors, which was expected to facilitate responsiveness. Five of these individuals declined due to personal reasons. The remaining 13 educators contributed to this investigation by completing individual interviews with us that lasted 1 h--1.5 h. All of the interviewees were leading quantum researchers at higher education institutions and had experience teaching quantum-related courses at their respective universities. Six educators worked at the same higher education institution, though not all were from the same department. The other seven worked at seven different institutions. Three educators taught at institutions outside the U.S., but all participating institutions were located in countries where the majority of the population is White. There were no qualitative differences between the U.S. and non-U.S. participants.
The educators were asked many questions, including their perspectives on how to diversify QIST \cite{ghimire2025reflections},  how to counter misinformation in the field \cite{kashyap2025strategies}, their experiences with QIST courses and curricula \cite{fargol}, their thoughts on issues related to the development of a framework for QIST education \cite{liamcommon} and their views on navigating hype, interdisciplinary collaboration as well as university-industry partnership in QIST \cite{hypeliam}. 

The interviews utilized a semistructured think-aloud protocol and were conducted via Zoom in a conversational manner, guided by an interview outline consisting of questions related to previously planned topics. While the interviewer followed this question list, the exact sequence of questions varied naturally depending on the flow of the conversation, and quantum educators were asked to explain and elaborate on their responses. All interviews were recorded and transcribed automatically. Transcription errors were corrected by listening to the recordings. Repeated words, such as ``you know,'' ``like,'' ``sort of,'' and other similar common filler words or phrases that participants used in conversations were removed from transcriptions for clarity. To preserve the participant's authenticity of responses as much as possible, we primarily present their answers in direct quotations to minimize the risk of misinterpretation and ensure clarity. Therefore, we only used paraphrasing for instances where the original quotes from the educators were challenging in terms of readability or comprehension.

In this paper, we focus on their response to the following critical questions related to QIST to get vetted information and a realistic timeline on important issues shaping the field:

\textbf{How would you respond to students, the public, and media who often ask the following questions?}

\begin{enumerate}[label=\textbf{Q\arabic*.}, leftmargin=*, itemsep=0.4em]
\item[Q1.] Do we have a quantum computer, considering we are still in the NISQ era? 

\item[Q2.] What is your estimate for how long it will take us to have a fault-tolerant quantum computer? 

\item[Q3.] Will we have a scalable quantum computer that can run Shor's algorithm~\cite{shor1994} with a large enough number of qubits to have a quantum advantage? If so, what is the estimated timeline for it? 

\item[Q4.] Even if you are optimistic about successfully building fault-tolerant scalable quantum computers, do you think we will ever have a quantum computer in our pocket?

\item[Q5.] What is your favorite qubit architecture that you think will play a major role in the future? Why?
\end{enumerate}

These questions are similar to those that not only one of the authors has been asked by their students in quantum courses, but also to those that other quantum educators have informally discussed with them that students in their quantum courses or in quantum outreach programs (to pre-college students) were asking these types of questions.

The researchers used structural coding for first cycle data analysis, similar to our earlier approaches to organizing interview data \cite{kashyap2025strategies,ghimire2025reflections}. Structural coding is a holistic, question-based method that labels and organizes responses according to research questions~\cite{saldana2021coding,hedlund2013overview}. This method of coding often results in the identification of larger segments of text (which we refer to as themes here) that are driven by specific research questions and helps in exploring commonalities and differences in responses of participants in the next steps of analysis. Once the data were organized in this way, we proceeded to do second cycle coding by identifying recurring ideas or patterns across participants' responses \cite{saldana2021coding}. This method helped us develop more focused categories within the structural code segments, resulting in grouping responses within each theme or research question based on similarity with other responses under that theme. We also ensured that if educator responses were nearly identical (i.e., restating essentially the same ideas of other educators), we only presented the response that best encapsulated the idea for conciseness. The second cycle coding was necessary to move beyond broader question-driven themes to both identify main ideas in quantum educator responses and improve comprehensibility for readers. At times, this resulted in codes with fewer responses from educators; however, these codes were retained because they represented ideas that added depth to the overall analysis based on multiple discussions among the researchers. Additionally, since some of the responses to the research questions asked were speculative in nature, some of the codes did not reflect a direct answer to the research questions; rather, they were focused on the educators' views and reasoning, as encouraged by the structure of the interviews.

\begin{figure*}
    \centering
    \includegraphics[width=0.8\textwidth]{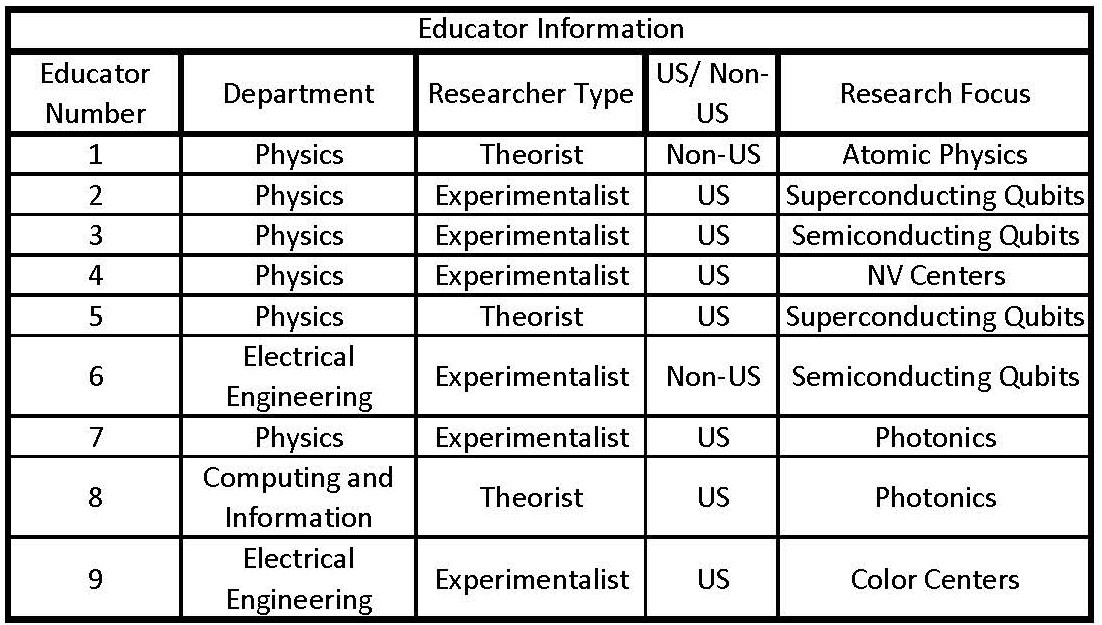}       
    \caption{This figure provides information on the educators, their departments, whether they are theorists or experimentalists, if they are based in the U.S. or not, and their most relevant research focus (some educators had other research focuses as well, not relevant to this work).}
    \label{fig:educators}
\end{figure*}

The coding of the data was conducted collaboratively, and the authors agreed on the codes after several rounds of discussion and refinement to reach final consensus. Although 13 quantum educators were initially interviewed, here we only focus on the responses of nine of them which directly and effectively addressed the research questions posed in this paper. We used a purposive sampling strategy to select responses \cite{patton2002qualitative}, and there were several reasons for not including specific educators in the results; for example, one of the reasons was the lack of relevant responses. The original interviews were lengthy and focused on multiple topics, and due to time constraints, not all of the participants addressed our research questions during the interviews. In other cases, the educators did not provide answers that aligned with the main research questions and were either too broad or highly specific in their responses. Therefore, the researchers agreed on including data that they thought conveyed the most important ideas, avoided repetition, and maintained clarity. The educator names (numbering) are arbitrary.
More information about the quantum educators, including their department information and research focus, can be found in Fig. \ref{fig:educators}.

\section{Results}

\begin{figure*}
    \centering
    \includegraphics[width=0.8\textwidth]{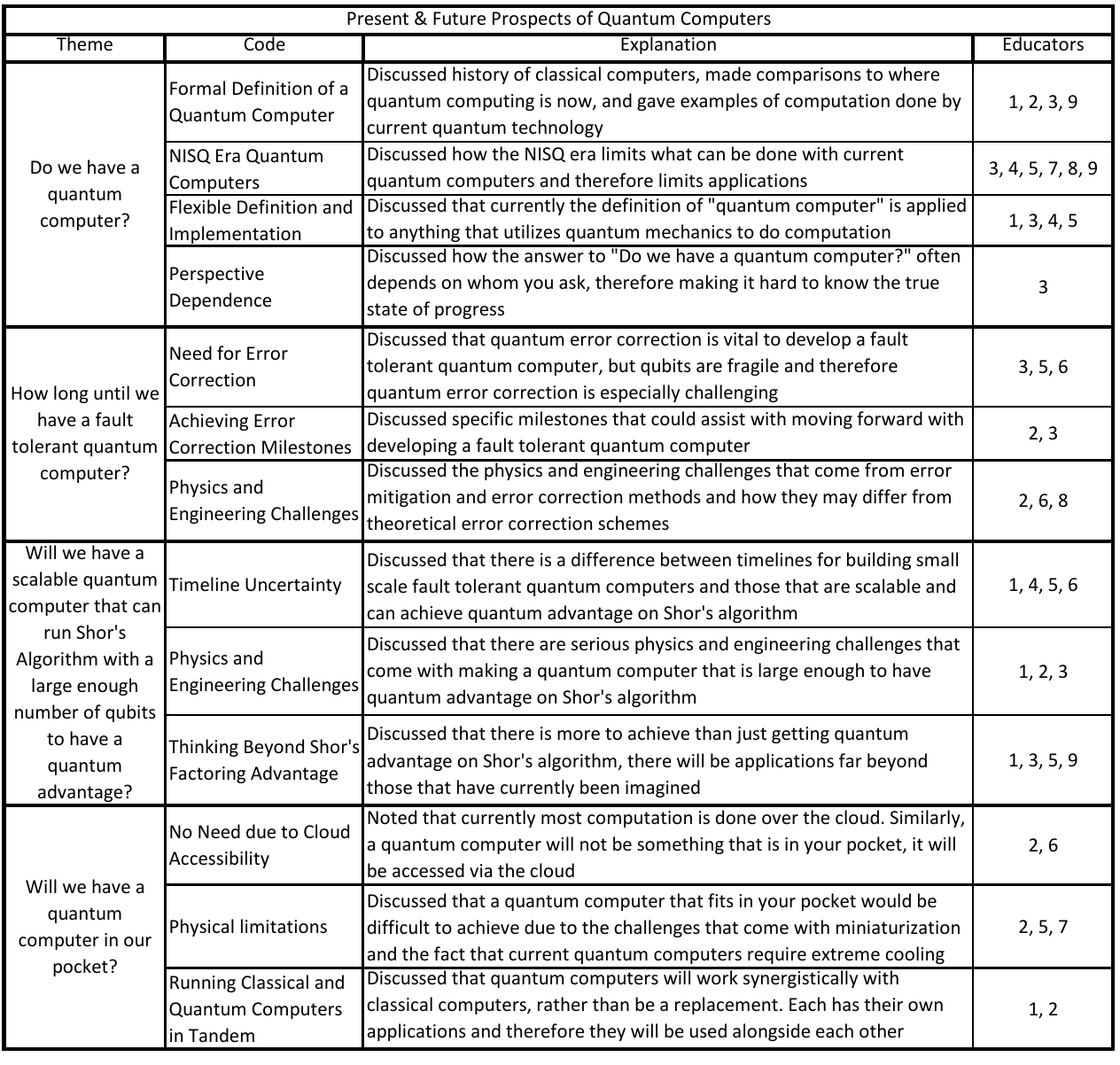}
    \caption{This table provides information about the themes (based upon the research questions),
    the codes for educator responses, explanations of the codes, and which educators contributed to the discussion relating to each code. }
    \label{fig:RQ1}
\end{figure*}

\begin{figure*}
    \centering
    \includegraphics[width=\textwidth]{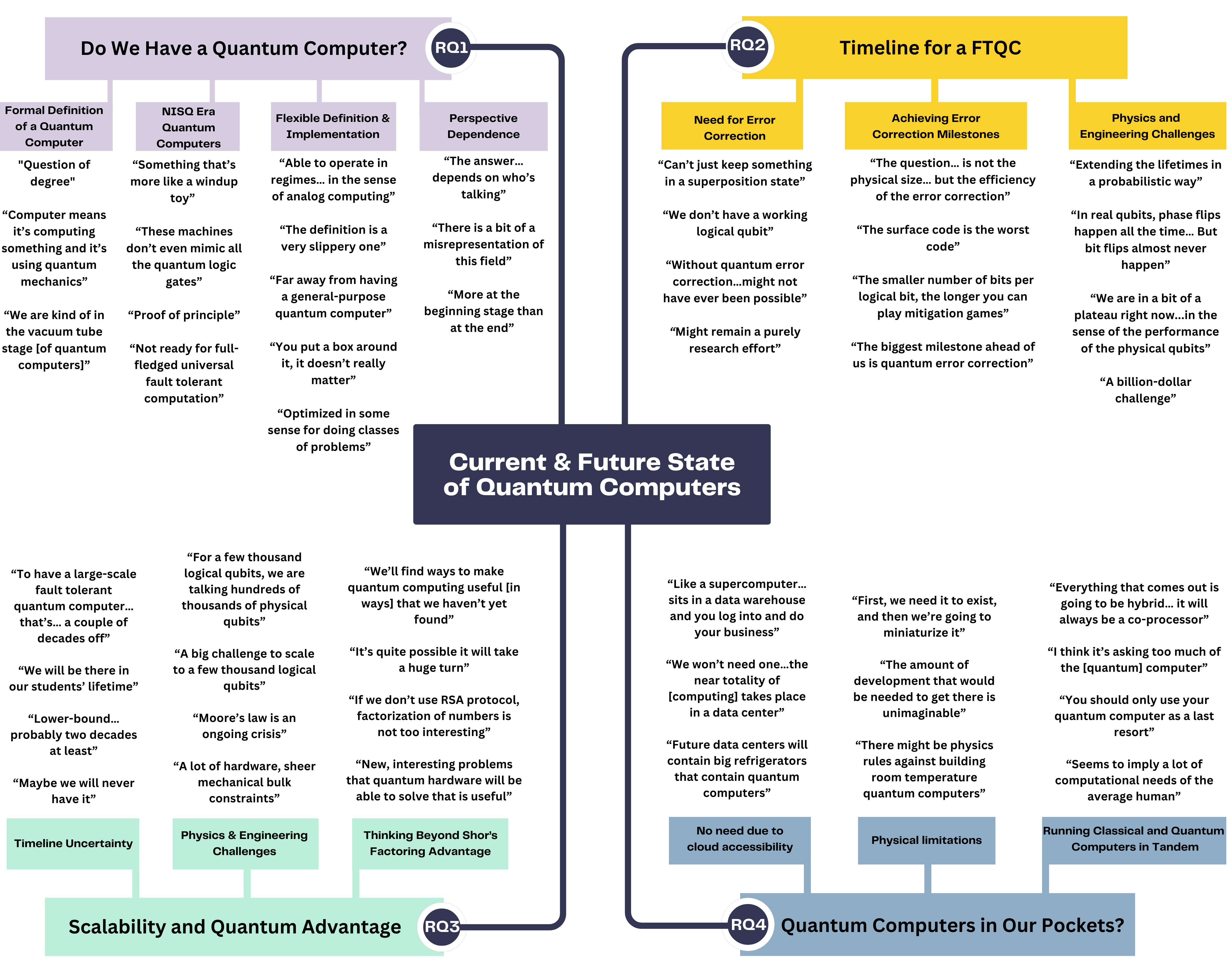}
    \caption{Examples of educator quotes on the research questions about current and future state of quantum computers, organized by recurring codes. In the figure, FTQC is an acronym used for fault-tolerant quantum computers.}
    \label{fig:summary}
\end{figure*}

Below, we discuss the findings organized by the themes based on the structural categorization corresponding to the research questions and the subsequent codes within each theme, before presenting an overall discussion of our findings, summary, and conclusions. Fig. \ref{fig:RQ1} describes the themes, codes, explanations, and which educators discussed each code. Additionally, Figure \ref{fig:summary} presents examples of educator statements under each research question and subsection to provide an overview of their perspectives.

\subsection{Q1: Do we have a quantum computer, considering we are still in the NISQ era?}

Quantum educators were consistently affirmative in their responses to this question. However, their responses were nuanced. Some of them focused on the definition of a computer and on what qualifies a machine to be called a quantum computer, drawing analogies with the gradual evolution of classical computers over the last century. They also noted that those doing research in quantum computing have become more flexible in how they define a quantum computer over time and emphasized that a quantum computer need not be circuit-based. The educators further stressed the importance of making broad investments across different qubit architectures at this early stage of development and noted that quantum computing is likely to show valuable results in quantum simulation in the near future. In addition, they pointed out that there is often a perspective dependence to what a quantum computer is, and that researchers are likely to have a different perspective, e.g., from investors and the public. The four codes for this section are as follows: formal definition of a quantum computer, NISQ-era quantum computers, flexible definition and implementation, and perspective dependence.

\subsubsection{Formal definition of a quantum computer}

Some educators focused on historical perspectives on how the word computer has been used. Some stressed that classical computers have evolved over time, but even the ones from a century ago should still be called a computer. By drawing an analogy, they emphasized that the definition of a quantum computer is a machine that performs computation using principles of quantum mechanics such as superposition, entanglement, and interference. 

For example, educator 1 reflected on whether we have a quantum computer, saying, ``I think that in the end these things are all questions of degree...It's a little bit like asking, in the early part of the 20th century, did we have computers? We certainly didn't have the digital computing architectures that we had by the end of the 20th century, but we had the precursors to all of these things, and they didn't have the advantages of fault-tolerant operation in many cases, they were not able to do the things that the computers were able to do at the end of the 20th century. But I would still say that we had computers. And if you go to the computer history museums that you find in various parts of the world, we will find machines from 120 years ago, and I would say that those were computers, and that they are the predecessors of things that we have, even though the architectures improved, and the capabilities were completely transformed. So yes, I would say, we do have a quantum computer.''

Educator 2 reflected on this issue, saying, ``I guess with new technology, there's always confusion because what does the word computer mean? It means...[something that] computes something.'' Educator 2 pointed out the historical parallels between classical and quantum computing, explaining that for early computing devices, it was not a question of how many transistors and what computational complexity or speed would make them be qualified as computers. Instead, they highlighted that the label computer was borrowed to describe devices with specific purposes during World War II: ``...to describe the thing that they had built, they borrowed this term: computer...That's why some of the original computing work was done at artillery development centers, because what they were computing was range tables.'' They continued, ``So, in the same sense, what is a quantum computer? I'm not aware that it has a particular computational [metric] that it can run X problem of X size before it's a computer or not. We're building these strange machines. We're trying to describe them to other people and just like before, we're borrowing this other term computer now double borrowed. And putting the word quantum on it, but it's not clear to me that even a scientist has a definition of what it is.''

Educator 3 focused on the meaning of the words ``computer'' and ``quantum'' saying, “...Computer means it's computing something and it's using quantum mechanics.” From this perspective, anything that uses quantum mechanics to do computation can be considered a quantum computer. At the same time, they emphasized, “But the rules of the game are different based on who's doing what. So, if you say, well, I want to minimize some Hamiltonian that I'm working on, and my system is quantum, and I can control [it], that's a really cool thing to do. But it's not the same as somebody who says, I want to create quantum logic and a sequence that can implement a quantum algorithm.”

Educator 9 said that the machines we have today are indeed quantum computers and expressed satisfaction that in the second quantum revolution, anyone who has access to the internet can access quantum computers from certain companies that provide free public access. They said, ``we are kind of in the vacuum tube stage [of quantum computers], but with the whole world having access to it. That was not the case with the first [classical] computers. They were in a room at an institution. If you were privileged to be at that institution, great, you could get training in it. And with all these books written about Bill Gates, why he was able to get ahead of competition, it is because his private school had a [classical] computer when he was young enough to make a difference.'' Therefore, they considered the accessibility of these otherwise expensive tools as a positive and unique feature, available to anyone with an Internet connection.

\subsubsection{NISQ-era quantum computers}

Several educators focused on the fact that the confusion regarding whether we have a quantum computer or not stems from the limited capabilities of NISQ era quantum computers and the fact that these NISQ machines are not capable of showing quantum advantage on problems that are of interest in real life. Also, different machines, e.g., with different qubit architectures, are often optimized for different applications, which further complicates how these machines are understood and classified.

For example, educator 3 clarified how NISQ-era quantum computers are different from what people originally envisioned, saying, “So, the kinds of quantum computers that people initially envisioned were ones that had the analog of quantum logic gates--so there's a circuit, they can basically implement a circuit model, which is similar to how classical computers work, but obviously different. And so, do we have a quantum computer? Well, I would say that we have something that's more like a windup toy. You have something that can clap like a cymbal and advance right across the floor. It acts, walks, and talks like a quantum computer. And then it stops. So, the point is that it doesn't get very far. You can wind it up only a certain amount and eventually it will stop.'' Educator 3 emphasized that the complexity of the actions that can be undertaken by a quantum computer is limited: ``...[many of these machines] don't even mimic all the quantum logic gates that are needed to do universal quantum computation.”

Educator 3 also reflected on issues related to error correction and fragility of qubits, saying, “And we haven't reached those milestones yet like quantum error correction. We don't have that. And if we didn't have ordinary error correction [on classical computers], there's no way we would be able to work with [classical] computers. So, we take all this for granted, we think, oh, yeah, I'm putting in a RAM stick or flash drive [and it works]. Many people don't appreciate that there's a lot of information technology that goes into making sure that the 1 is really a 1 and doesn't flip to a 0 [in a classical computer].''
They continued, ``And it's worse with quantum bits. We know that quantum bits lose their information over time... [but] does that mean it's not a quantum computer?” Educator 3 explained that despite the issues involving computation in the NISQ era, the current machines running toy models for computation are still using quantum mechanics to do computation, so they are, in fact, quantum computers. However, these nuances should be made clear to those asking the question of whether we have a quantum computer or not.

Educator 4 acknowledged the limitations of NISQ-era machines, e.g., for leveraging the power of quantum algorithms \cite{shor1994,grover,vaziranib,simon}. They said, ``let me put it this way. The tasks that [a quantum] computer can do at the moment are completely simulatable on your laptop or your desktop computer. But does that mean they're not doing quantum stuff? No, it doesn't. They are doing quantum stuff. It's just that the power of the quantum algorithm cannot be leveraged yet. There are superpositions that are entangled. There is interference going on in these quantum computers as they execute the algorithm. But the scale of the algorithm, because of the number of qubits...in this noisy regime is not enough to give us the power, the exponentials, the super polynomial speed up that we would like to see from quantum computers.'' Educator 4 emphasized that current machines are not universal quantum computers. They are not error corrected or fault tolerant, and significant progress is needed to get there. They asked, ``But are we executing quantum algorithms? Are we creating entangled states...Let's be pretty blunt. This has not been done before. There's no doubt about that...if you take something like the sampling circuits and the sampling algorithms for boson sampling distributions. We don't strictly speaking care what happens in those, but it's a clear demonstration that you are able to achieve something that a supercomputer would not be able to achieve...So, I would say that we have quantum computers.''

Educator 5 said that current quantum computers in the NISQ era can be thought of as proof of principle, saying, ``I will probably say that we have really small and imperfect quantum computers right now that may be proof of principle. We can do better, and we are on the cusp of making them kind of do useful things for certain tasks...But if you want to think about an actual, fully fledged, universal quantum computer in analogy to a classical computer you can buy today, that's still decades away.''

Educator 7 had similar thoughts and brought up examples of why car companies have researchers accessing certain types of quantum computers in the NISQ era. They emphasized that currently, ``there's no error correction. So, I think it's more research that they [those using the current quantum computers] are doing. [For example,] companies like Honda...the auto companies, they're renting time on these NISQ computers basically to do research in quantum computing so they can try to develop algorithms and test them in some very rudimentary way, in part, so these companies will train there [on these quantum computers]. They hire scientists who are doing quantum information, quantum computing, looking to the future. Now they're trying to do things like a better battery design. And they claim that they've already been using IonQ \cite{ionq} quantum computer [a company that focuses on trapped ion quantum computing] to simulate some chemical reactions...It's probably not circuit-based. It's probably simulation.'' Educator 7 added, ``most of the things being done are simulation, although the one that Google did, when they sampled from a random number generator, that was circuit based. But then other people said, there was no quantum advantage really in that case.''

Educator 8 also acknowledged the limitations of the current machines, saying, ``they [NISQ-era machines] are quantum computers, but they are not ready for full-fledged universal fault-tolerant quantum computation”.

Educator 9 was extremely positive about the recent progress in quantum computing and noted that the existing machines in the NISQ era are quantum computers. They focused on the range of applications already being pursued with quantum computers, saying, ``I think there are so many groups that are working on using the current [quantum] computers for something that could be useful for them. In our case, we are using them to simulate and guide light matter interaction. And I think it's impressive, just from an engineering point of view and physics point of view, that humankind has been able to control entanglement across so many qubits spatially and temporally. It's fantastic for something that usually occurs in such exotic situations, low temperatures, or small sizes [to be controllable and manipulatable the way we are able to do]. And while there are so many challenges that we are all addressing in our respective disciplines, progress has been relatively quick.''

\subsubsection{Flexible definition and implementation}

Some educators acknowledged adopting a flexible definition of what a quantum computer is, suggesting that any machine that uses 
quantum mechanics to do computation can be called a quantum computer, regardless of whether it is circuit-based or not. Some educators also emphasized that over time, the research community has generally become more flexible regarding what is considered a quantum computer. Additionally, many educators recognized that the current quantum computers will primarily have their usefulness showcased in quantum simulations in the near future.

For example, educator 1 said, “…my feeling is that we do have devices now that are able to operate in regimes where they can reliably, quantitatively do calculations in some sense, and at least in the sense of analog computing. You build a model. You make quantitative measurements that certainly give you solutions to that model in regimes that we don't have access to via this conventional classical computing techniques. And for that reason, I would say, yes, we do have quantum computers, and we do have quantum simulators. And at the moment, they're running in analog regimes...But then I would be very careful about saying that these are quantum computers that are [only] able to, at the moment, solve some specialist problems that are of interest to researchers, at least in physics. We might be close to making them of interest to people in materials science.''

Educator 3 contemplated these issues and acknowledged that, ``the problem is that the definition [of a quantum computer] is a very slippery one, and people have adopted or co-opted...to call whatever the hardware they have a computer. But it really depends on the definition”. Educator 3 emphasized the need to think big picture and be expansive saying, ``the kinds of examples that I was giving, they [quantum computers] can solve certain models and do certain things...they're kind of toy problems in a sense. So, the question is, how do we decide what [types of quantum computers] to develop?'' Educator 3 explained that while some may only focus on circuit-based quantum computers, others may be more interested in simulation, particularly for advancing fields such as material science. Therefore, educator 3 believed that it is necessary to be open to the development of quantum technologies to solve different types of problems. ``We don't even know whether they could end up interacting together and helping one another in some future architecture. You put a box around it [a quantum computer], it doesn't really matter [whether it is circuit-based or not], because [the only important thing is that] it is solving a problem.''

Educator 3 emphasized, ``We're pretty far away from having a general-purpose quantum computer. I just don't think that's gonna happen anytime soon. But it may be that there are more specialized quantum computers that are optimized in some sense for doing classes of problems. And there I can imagine solving quantum chemistry problems or even optimization problems...I'm not sure that they're gonna look exactly like textbook circuit problems that are kind of well established.'' 
They also noted that while we are using quantum computers for toy problems that are currently doable, theoretically, any quantum Hamiltonian can be turned into a quantum circuit involving a collection of qubits and quantum gates \cite{lloyd,simulationfeynman} and it can simulate time evolution of a quantum system. They said, ``So this is the thing that we know is possible with a circuit-based quantum computer. Seth Lloyd took to heart what Richard Feynman said about quantum computers ...if we could control matter, then we could essentially solve the Schr\"{o}dinger equation and learn a lot about the quantum nature of our universe. Seth Lloyd...showed that you could take any Hamiltonian, and you could write down and turn it into a bunch of qubits and gates and simulate the Schr\"{o}dinger equation. You could write down the time evolution of a given state.''

Educator 4 also stressed the importance of approaching the concept of quantum computers with flexibility, pointing out that quantum simulations will become useful before we can expect quantum computers to have a quantum advantage \cite{advantage} on solving problems such as implementing Shor's factoring algorithm. With an eye toward applications in quantum simulation, they said, ``When people say climate change and quantum computers, what are they talking about? That quantum computer could be used to simulate some new materials and thereby help in solving climate change problems, for example, some material that will do carbon capture better or better battery material...And I think the same is true for many experiments. I know often, for example, when I visit say QuEra \cite{quera} in Boston, they are focused mainly on things like that, trying to make arrays of atoms and make them into a quantum simulator with a thousand qubits, for example. Even D-Wave \cite{dwave}, for example, when they do their annealing stuff, they're really thinking of using them as a way to optimize...I'm sure that in Google and IBM business plans, this is probably the first thing that's on the horizon, trying to use these [quantum] computers for simulations of other problems, whether there'd be some optimization problem or maybe some variational algorithm for simulating drug molecules or whatever it is. I think that's really what most people in these NISQ computers recognize will be the immediate applications.''

Educator 5 similarly acknowledged that over time, the definition of a quantum computer has become more flexible. ``So if we go back 20 years, what people will call a quantum computer, what they really meant was a universal quantum computer, which is something that you can program, say, execute...Shor's algorithm on. For sure, we don't have a scalable one yet...Now we have small universal quantum computers...you can only run so much before the system decoheres. So, in that sense, it kind of exists...And then there are other...quantum computers...where they would do some quantum tasks reasonably well, but you can't program it to do any calculation. So, for example, quantum annealers and then various applications of this are definitely used in quantum information technology...they are trying to do certain tasks with better sensitivity than existing systems and trying to beat some classical computer in some tasks, the so-called quantum supremacy [also called quantum advantage]. So, it depends on what you really mean by a quantum computer.''

\subsubsection{Perspective dependence}

Different stakeholders, e.g., researchers, investors, and the public, can have different understandings of what a quantum computer is, causing further confusion about the current and future state of quantum computing. For example, venture capitalists, investors, and the public often hear about advances in the number of physical qubits in a quantum computer with a certain qubit architecture, not realizing that we have not yet achieved a fault-tolerant quantum computer of any size, let alone a scalable fault-tolerant quantum computer that can give quantum advantage, e.g., on Shor’s algorithm. Although only one educator elaborated extensively on this viewpoint, four other educators 
noted at some point in the interview (in response to other questions) that business leaders and/or venture capitalists tend to exaggerate the progress made in quantum technologies, which can impact these issues and obfuscate the timeline. We did not include those responses explicitly because they were closely tied to other research questions, which will be discussed in future work.
Yet, we decided to emphasize valuable insights provided by educator 3 as a separate code, which summarized these points that the other educators also made.

Emphasizing diverse definitions of a quantum computer for different stakeholders, Educator 3 said, “I would say that the answer [to the question of do we have a quantum computer]...depends on who's talking. When you start to throw investors and the public into this next, it becomes even more confusing...[They think that] this is going to be so cool and we're going to make a lot of money, and then you realize that, okay, maybe it wasn't living up to expectations. But again, they were referring to this [e.g., a scalable fault-tolerant quantum computer as a quantum computer], and you thought it was that [toy models we currently have in NISQ era]. So, there was a misunderstanding about what it is. So, all of that is another dimension altogether [related to the issue of do we have a quantum computer]...''

Educator 3 also stressed that the confusion about whether we have a quantum computer partly stems from the fact that we are at the early stages of the development of quantum technologies. However, many stakeholders [including investors and the public] who are not researchers have an incorrect view of our current stance in this progression, partly due to the way information about these issues is presented and consumed. They said, ``I think there's a bit of a misrepresentation of this field. It's usually described or depicted as a mature technology, and it is anything but. We're working at the beginning of something, not at the end.'' They made an analogy with the first transistor, saying, ``So, if you've ever seen a picture of the 1st transistor, it's this big messy thing, with a triangle hitting another surface. It ended up being the iconic representation of a diode. But the point is, there're wires hanging out. And it looks nothing like the nanoscale transistors that we have today.''
They continued, ``We are sort of at that stage [in this quantum computing field] where we're trying to work with things that are relatively messy, inefficient. They don't work well. We have an idea of what we want, but we don't really know the pathway to get there. We can look at a shiny dilution refrigerator with thousands of coaxial cables, and it looks like this amazing thing. But actually, this is just the fridge, it's not the quantum part of it. And so, the real challenges are out in front of us. So, we're actually more at the beginning stage than at the end. And the beginning stage is more defined by really basic science and engineering stuff, these are things that we are trying to deal with [right now], things that would be impactful if we can reach certain milestones.”

\subsection{Q2: What is your estimate for how long it will take us to have a fault-tolerant quantum computer?}

Most quantum educators believed that within 10 years, we would have small, fault-tolerant quantum computers \cite{quera,ibmscalable,googlescalable}. Their optimism came from the expectation that quantum error correction or quantum error mitigation would be possible in that time frame for making a quantum computer with a few logical qubits. They emphasized that achieving quantum error correction milestones in any qubit architecture is currently very challenging. However, continuous investment is very important in order to overcome the physics and engineering challenges. Due to the educators' agreement regarding a small, fault-tolerant quantum computer being achievable within approximately 10 years, the codes for this section instead relate to the specific requirements that the educators emphasized when discussing how this timeline would be met. The three codes for this section are as follows: need for error correction, achieving error correction milestones, and physics and engineering challenges.

\subsubsection{Need for error correction}

Educators emphasized the importance of quantum error correction for building fault-tolerant quantum computers. They stressed the fact that qubits are fragile and decohere easily, and that quantum error correction is challenging. Thus, figuring out the errors that have emerged in the qubits during computation and correcting them is a very challenging task.

For example, educator 3 brought up the need for error correction, saying, ``You can't just keep something in a superposition state and expect it to remain in that state indefinitely. No quantum bit works that way right now.'' They acknowledged the possibility of error correction and improving a qubit, but that we have not reached that point yet. ``We're kind of at the break-even point, maybe in some [qubit] platforms. But we don't have a working logical qubit that's capable of storing and maintaining and processing quantum information with a manageably small amount of errors. And so, this is one big obstacle that stands in the way of actually using quantum computers."

Educator 5 believed that Shor's contribution to quantum error correction \cite{shorcode}, around the same time he came up with Shor's factoring algorithm, played a key role in the possibility of building a fault-tolerant quantum computer. They said, ``so Shor is actually responsible for 2 key innovations...and they're basically back to back. If it was just Shor's factoring algorithm and nothing else, it [QIST] might still have not taken off, because without quantum error correction, it might not have ever been possible to build a machine with the kind of fidelity where you can actually run a quantum algorithm to any usefulness. It took Shor's work [on error correction], plus a few follow-ups for people to start believing that we can actually do so [error correction].'' 
They noted that even if we do not have robust logical qubits now, they were optimistic that error correction would be possible going forward. 
``I think the other thing [that Shor did], error correction, might not be given enough credit...there are also all the technologies that [were developed] at the same time, for example, that was the time when superconducting [qubit] technology started taking off and stuff like that. So, there were a number of things that kind of happened at the same time...when people started taking it [prospect of building fault-tolerant quantum computers] seriously [in the mid-nineties].'' Educator 5 also said that they were hoping we would have small fault tolerant quantum computers with error-corrected qubits in 10 years but felt that quantum computing still ``might remain a purely research effort'' at that time.

Educator 6 was reasonably optimistic about the prospect of being able to successfully do error correction and build at least a small fault-tolerant quantum computer. They said, ``Predicting the future is notoriously hard. I think it will take another 10 years before we have something that really makes a difference."

\subsubsection{Achieving error correction milestones}

Some educators discussed how quantum error correction or quantum error mitigation can be achieved in specific qubit architectures despite all the challenges involved in quantum error correction. Additionally, some talked about using quantum error mitigation techniques to get the best out of noisy qubits and be able to do quantum computation.  Educators also emphasized that ensuring that we continue to make progress in quantum technologies with an eye toward fault-tolerant quantum computers is so important that we must continue to adequately invest resources to make it happen, despite the formidable hurdles. 

For example, educator 2, who works on quantum computing with superconducting qubits, said that their estimation for moving beyond the NISQ era and having small fault-tolerant quantum computers is a decade. They contemplated the issues involved in building a fault-tolerant quantum computer, saying, ``Really, the question there is not the physical size [e.g., of superconducting qubits], but the efficiency of the error correction. That's why I've really been focused on that. Again, if it's 10,000 [physical] qubits per [logical] qubit, everything is really terrible, though IBM has a roadmap that goes past I think 100,000 qubits. So maybe even there, they get there the hard way.” 

Regarding how they are planning to make logical qubits from physical qubits experimentally for superconducting qubits, educator 2 started by pointing out that a leading quantum error correction code, the surface code \cite{surfacecode}, is not good. They said, “They're not going to be the kind of things where it's 100,000 [physical] qubits that makes 10 [logical] qubits. But the one reason our correction always looks so crazy far away is because the surface code is the worst code. So, the reason I'm feeling optimistic is because I see in the literature and in talks, people starting to find ways around the surface code where you don't need thousands of qubits per [logical] qubit.'' Educator 2 also pointed to quantum error mitigation, saying, ``The smaller is the number of bits per logical bit, the longer you can play mitigation games...I just know that one of the obstacles in the field is that a lot of these computer science derived schemes for error correction, and the algorithms and all this stuff are worked out in the infinite qubit limit. And that's not the same as the few qubits limit. So, I see people trying to find things that work well with a few qubits or maybe, for instance, that aren't infinitely scalable but work to a level that they can then build and concatenate. That to me looks like a big improvement that makes it feel not so impossible, because I probably don't need a million qubits [per logical qubit].''

Educator 3 stressed that advances in quantum technologies are so important that we should be thinking big, ``the biggest milestone ahead of us is quantum error correction...Will we be there [have error corrected qubits and be able to build fault-tolerant quantum computers] in 10 years? I don't know...There have been a lot of moonshot projects and problems. And I think only some of them happened in a timescale that we would have been comfortable with like a decade.” Despite this uncertainty, educator 3 strongly emphasized that we should continue to invest heavily in quantum technologies, since they are likely to revolutionize many aspects of our lives in ways that we have not even imagined.

\subsubsection{Physics and engineering challenges}

Educators discussed the physics and engineering challenges involved in quantum error correction and quantum error mitigation across different qubit architectures, with the goal of extending computation time beyond decoherence time. They also emphasized the importance of distinguishing between theoretical error correction schemes, e.g., those that may only be relevant for the large N (number of qubits) limit instead of systems with a finite number of qubits, or theoretical approaches that may emphasize both bit flip and phase flip errors instead of only the phase flip errors relevant in certain systems. They also stressed the importance of ensuring that for quantum computers with different qubit architectures, shortcuts are devised to deal with experimental challenges in quantum error correction in finite-size systems or approaches are considered that shortcut quantum error correction or mitigation in systems in which both bit flip and phase flip errors may not be equally important.

For example, to deal with physics and engineering challenges, educator 2 described one type of error correction scheme for superconducting qubits, saying, ``...[some] schemes for error correction are sort of extending the lifetimes in a probabilistic way. So, taking kind of bad qubits and building subspaces inside them so that you can try to keep [the qubit] alive for much longer. So, if [one] can make that work, then 10 years is enough for 3 or 4 generations of experiments. We might have some machines with 10 [logical] qubits in them.” For them, the central question was how many different shortcuts and methods can be found for quantum error correction or mitigation for their superconducting qubits
so that they can take advantage of them for real physical challenges and not those from a purely theoretical or computer science perspective. 
``So, for instance, in real qubits, phase flips happen all the time like T2 [time]... But bit flips almost never happen. A bit flip is like someone comes in and does an unintended pi pulse on your qubit over the x or y axis...So, I'm optimistic that if we can really figure out how to build schemes that work on what we have in real life and not based on sort of too idealized abstractions, we can shortcut this, and then lots of things have a chance to scale.''

Educator 6 also reflected on physics and engineering challenges involved in building a fault-tolerant scalable quantum computer and said, ``that's very, very hard. But these things are not linear. We are in a bit of a plateau right now I think...in the sense of the performance of the physical qubits. So, there has been in the last 5 to 10 years a big push towards making more and more qubits and making them work together. And that's been very successful. But the actual quality of the individual physical qubits has improved, but not hugely so. So there needs to be another 2 or 3 orders of magnitude improvement in that.'' They emphasized the critical role of academics in driving this improvement: ``That's why I think there's still lots of space for academics to play a role. Even though I will not build the fault-tolerant [scalable] quantum computer in my lab, 
you're in the university, it's just impossible. I will never have the resources to do that. That's a billion-dollar challenge. I don't have that [money]. But I have the creativity. I have the understanding of microscopic physics to help push the quality of the physical qubit, for example, and that then can feed into the large-scale quantum computers built in the industry.''

Educator 8 was also optimistic that we will be able to overcome the physics and engineering challenges, but did not provide a timeline for a fault-tolerant quantum computer. They pointed out that we have physical qubits with no error correction at this time, and that researchers are currently trying to demonstrate logical qubits. They continued,
``But then, to start doing computation with these logical qubits, it's going to be a huge engineering challenge. But I think there is a lot of promise from what we hear in the industry and also in academia.''

\subsection{Q3: Would we have a scalable quantum computer that can run Shor's algorithm with a large enough number of qubits to have a quantum advantage? If so, what is the estimated timeline for it? }

Overall, most quantum educators were hopeful that a scalable quantum computer with quantum advantage would eventually be achieved, but their estimated timelines showed some variations. 
It was also pointed out that even after achieving this goal, having a quantum advantage for Shor's factoring algorithm is not going to be a useful goal. Instead, we would certainly discover many applications of quantum technologies that are currently beyond our imagination. The three codes for this section are timeline uncertainty, physics and engineering challenges, and thinking beyond Shor's factoring showing quantum advantage.

\subsubsection{Timeline uncertainty}

While many educators predicted that we would have small fault-tolerant quantum computers within 10 years, their predictions for achieving quantum advantage on Shor’s algorithm reflected greater uncertainty. For example, one educator put a lower bound on the latter timeline, saying it probably will not happen for at least two decades, while another educator expressed excitement about the possibility of ultimately finding out that some law of nature may prohibit the realization of a scalable fault-tolerant quantum computer with quantum advantage on Shor’s algorithm.

For example, educator 1 appreciated the possibility of passing  
essentially all the thresholds of fault tolerance \cite{errorcorrection1,errorcorrection2,surfacecode,kitaev} and the demonstrated potential to scale to arbitrary fault tolerance. They stated, ``I think that will happen on a small scale in the next couple of years, and that we will have that relatively soon. But to have a large-scale fault-tolerant quantum computer of the scale that would be able to run Shor's algorithm and factor a substantial size number that you can't factor on a classical computer at the moment, that's going to be another few steps. There's going to be a lot of challenges in the physics and engineering that probably [will] require different details [for] some of the architectures. So that's going to be probably a few, could be couple of decades off.”

Educator 4 expressed optimism regarding building a scalable quantum computer that can provide computational advantage on Shor's factoring algorithm, saying they hoped that at least ``we will be there in our students' lifetime.''

Educator 5 predicted a lower bound for this timeline saying, ``I don't have an upper bound for when we'll definitely get it. But I have a lower bound for when we won't get that, and that is probably two decades at least before we get to where
..it can solve a problem that so far, no classical computer has solved [with practical usage].'' 

Educator 5 made an analogy between classical and quantum computers' timelines saying,
``...the famous Moore's law, which is [that] the number of bits [in a classical computer] doubles every 18 months...if [for quantum computers], you try doing a slope similar to Moore's law, it's a much...shallower slope, which kind of means that even optimistically, if you think some...law [similar to Moore's law for classical computing] applies to quantum computing, its time scale will be much longer in comparison...it's gonna take even longer for this one. The transition from vacuum tube to semiconductor [transistors that revolutionized classical computing] to miniaturization era, in retrospect, you knew it was coming. But when you were in that era, it took a long time. And for quantum computers, it's going to go through hopefully that process, but it will definitely take longer [than the time for classical computing architectures to mature].'' Regarding the estimated timeline for quantum advantage on Shor's factoring algorithm, educator 5 added, ``perhaps...I will see it in my lifetime but who knows.''

Educator 6 remained open to all types of scenarios, including the possibility of never succeeding in building a fault-tolerant scalable quantum computer. They said, ``...maybe we will never have it. I'm at peace with that. To me, in a sense, one of the most exciting things that could happen is that we will discover a new law of nature that prohibits fault tolerant [scalable quantum computers that can give computational advantage on Shor's algorithm over classical computers] and I think it would be the most exciting thing, more exciting than finding the Higgs bosons or gravitational waves.''

\subsubsection{Physics and engineering challenges}

Some educators explicitly pointed to the inherent physics and engineering challenges that will need to be overcome to gain quantum advantage on Shor’s factoring algorithm. They emphasized that, at this early stage of quantum technologies, it is crucial to continue to embrace and persevere with these challenges, and strive to optimize systems with all types of qubit architectures in order to make progress.

For example, educator 1, who works closely with experimentalists and has recently been doing theoretical work on neutral atoms, shed light on these issues by giving an example of challenges in scaling to $\sim$1000 logical qubits. They said, ``It's very clear how you get these things [neutral atoms] to a few thousand physical qubits within the next couple of years...Of course, for a few thousand logical qubits, we are talking hundreds of thousands of physical qubits. There are things that need to be sorted out with the architecture of exactly how you do read out of that sheer number of qubits, and how you would do the engineering for essentially all of the lasers addressing all of those qubits.'' They continued to make a comparison with solid-state cryogenics systems, saying, ``But the problem is not dissimilar of course to the problem that you have in any of the solid-state cryogenics systems, where essentially you need all of the different electrical control lines corresponding to the different qubits in some form. So, it becomes a very similar but different engineering challenge.'' They expressed optimism about achieving working logical qubits, but that it comes with its engineering challenges:
``It is a big challenge to scale to a few thousand logical qubits, because of the hundreds of thousands physical qubits that you need to control. But I think that the scale of the challenge is not too much different for neutral atoms than it is for the other leading platforms, and there are elements of it that will be easier.''

Educator 2 reflected on the physics and engineering challenges involved in making sure that classical computers continue to follow Moore’s law, and how challenging it will be to build a scalable quantum computer. They said, ``...Moore's law [for classical computers] is an ongoing crisis, and they keep completely changing out what the transistors are built out of to stay on Moore's law. Does that mean that we're gonna go from ion traps to neutral atom to superconducting qubits to new superconducting qubits to this to that? I don't know, though we as computer users don't really know or care about all the insane amount of rejiggering that happens in a computer to keep them moving forward. We're just happy that it is scalable''.

Focusing on the physics and engineering challenges, educator 3 said, “There are a lot of people thinking about these problems from various angles but there are a lot of hardware, sheer mechanical bulk constraints about how many wires you can fit down into a fridge and cooling power of the fridge that are really causing some problems with companies that are trying to develop scalable quantum computers. And again, we don't have a single protected qubit [in the current quantum computers].''

\subsubsection{Thinking beyond Shor’s factoring quantum advantage}

Some educators felt that we should be thinking beyond the quantum advantage of Shor’s factoring algorithm, as there are bound to be exciting applications of quantum technologies that we have not even thought about so far. They felt that quantum technologies are very promising and destined to lead to transformative applications in the future. Therefore, major investments in these technologies should be continued regardless of whether a quantum advantage on Shor’s factoring algorithm is reached or not.

For example, reflecting on the future of quantum computing with computational advantage in factoring products of large prime numbers on a quantum computer and beyond, educator 1 said, ``We've got quite a long way to go before we make them of interest to people who want to factor numbers. And...by the time we get to the point that we will factor numbers, most of the people who are interested in factoring numbers, will be using different ways to do their cryptography. So [they will not be interested in factoring] the numbers anymore...'' They were optimistic about achieving a scalable quantum computer with quantum advantage on Shor's algorithm, but their optimism went further: ``We will, on the way there, find much more interesting things to do with them than we've come up with up till now, and I think we'll find ways to make quantum computing useful [in ways] that we haven't yet found. The bigger we build the systems, the more people are gonna be trying to do crazy things on them and the more crazy new ideas you're going to get. So it may be that in 40 years time, [if] someone runs Shor's algorithm, it may be more of a curiosity than a practical application.''

Educator 1 also reflected on the broader future of computing, including the healthy race between classical and quantum algorithms inspired by potential classical approaches to thwart the quantum advantage. They said, ``I see a couple of different things. There may be ways to solve certain types of classical problems [on quantum computers], and I think that we should keep pushing that.'' They pointed out that researchers focus on different topics, such as partial differential equations or optimization, but it is difficult to predict whether they are going to see any advantage over classical techniques in those directions. Educator 1 continued,
``In fact, there is a beautiful race almost between people coming up with new ways to do things on the quantum side and people coming up with ways to do better than that on the classical side. So, it's also pushing the development of classical algorithms in a very positive direction and causing people to be creative in a way that they may not have been, because they see how you might do something on a quantum computer and [they] ask, `is there a different way to do things classically?' and that's very exciting.”

Educator 1 also emphasized the importance of quantum simulations and sensing to humankind. They felt that even if a scalable quantum computer achieving quantum advantage for Shor’s factoring algorithm will not happen in the coming decades, the benefits of quantum technologies would still be immense and that is what should be kept in mind. They said that as we move forward in quantum science, broader audiences will become interested in quantum mechanical problems. Educator 1 provided an example saying, 
``we say, well, this is quantum simulation. It might be of interest to a physicist now. But I think it'll be of interest to material scientists in the next few years. It'll be of interest to a quantum chemist in a slightly longer time scale...And even if it is just quantum simulation and maybe the use of these things for quantum metrology, then the impact that [these things] could have on society could be really quite huge.'' They continued, ``Don't forget or underestimate quantum measurement and sensing; because if you can build these big, entangled states that are for quantum computing, then you can start trying to use those to build better measurement devices. That means better sensors to do the next generation of fundamental physics and understand the universe. That means other potentials for new technologies. But again, I think also that a lot of these quantum mechanics-based subjects, material science and quantum chemistry, are themselves going to have a big impact on other areas of the real world outside of science. And so, I think we will find ways to translate the applications of quantum computing also to the wider world, even if the main focus of where we get advantages is in quantum simulation.''

Educator 3 echoed sentiments similar to other educators about why we should continue to invest in quantum technologies regardless of whether we get quantum advantage on Shor’s algorithm in the next few decades. They compared the pursuit of quantum technologies to past moonshot projects, some of which materialized quickly, while others have taken a long time to yield results \cite{ligo2013}. They said, “like saying we're gonna go to the moon and then landing on the moon in 1969. Maybe there will be successful fusion reactors even though they've taken a long time. If we think about being able to measure gravitational waves, it was failure after failure after failure until it failed to fail.'' They viewed advancing quantum technologies as both essential and fundamental, saying,
``We shouldn't be worried about whether it's gonna happen in this fiscal year or next fiscal year or this decade. It's something that we have to do. It's the most fundamental change to our way of doing computation since the development of [classical] computers and transistors.''

Educator 3 also emphasized expansive thinking beyond quantum advantage on Shor’s factoring algorithm, stressing that since predicting the future is challenging, we do not know what the truly transformative applications of quantum technology will be. They highlighted that the broader perspective is critical for harnessing quantum technologies’ full potential, saying, "I would say that because we're at the beginning of the second quantum revolution, we actually don't really appreciate what are the most impactful applications. We thought that when the transistor was invented, the biggest, most important thing was gonna be better hearing aids…be able to put something in your ear with a transistor. We were wrong, but it's because it's very hard to see far into the future. But we do know that this theory [quantum] is extremely important; it is the operating system of our universe, and we have unprecedented control over quantum systems. And we're learning how to use them to solve really hard problems, but which problems? It may be that we're still at that hearing aid level of things where we haven't really had the vision to see far into the future."

Educator 5 was reluctant to predict the future or whether and when we would have quantum advantage on Shor’s factoring algorithm. They said, ``so first, it's gonna be notoriously difficult to predict what happens ...[but] quantum technology, at least the current incarnation, is certainly not going to go away, I'm pretty sure. But I don't know [if] it's going to be packaged in the same way...'' Educator 5 also made an analogy with the evolution of classical computers and how predictions were far from reality: ``We're in the vacuum tube technology...but it's quite possible it will take a huge turn...If you go to the street and...ask people, what's the big thing back in 1920s, they'll almost certainly say things that in retrospect are ridiculous...If you look at history of what people hype about, for example...[there was] the hype about space age [and] there were plenty of things in every decade that people were excited about that did not pan out...and usually when we write history, we don't write out all the things people were excited about that did not pan out...probably 90\% of the things are hype... [that] did not pan out just because that's usually how it goes.'' 

Regarding other quantum algorithms for classical applications like Shor's algorithm for factoring, educator 5 noted, ``People have been struggling to kind of find cool algorithms that...have real-world applications. They have some that might have more restricted kind of research-based applications. But in general, it has been difficult, and there have not been many instances where you can say, oh, this algorithm on a quantum computer would definitely be cool because it can do xyz that a classical computer definitely can't do.'' Although educator 5 acknowledged the challenges in making fault tolerant scalable quantum computers, they were excited about the recent advances in quantum sensing saying, ``sensing technology, which isn't exactly computing, but it definitely uses a lot of tools that were developed for quantum computing, which is the ability to sense small signals...that is definitely one technology that I think has come on the horizon.''

Educator 9 also accentuated thinking beyond quantum advantage on Shor's factoring algorithm. They said, ``I wouldn't say that I'm 100\% sure that we'll have that machine and that advantage. Maybe we'll have that machine, but it turns out that it will cost so much energy or money or something to crack anything that it might not be useful like that. Maybe it will be used on a few examples and not really on an everyday basis because that was the idea for decryption.'' They believed that it is likely that by the time such a machine is ready, the world will have moved away from is RSA protocol, and there will be no need for factorization anymore. They said, ``If we don't have RSA, factorization of numbers is not too interesting. So that's why for me, I'm not following when we're gonna get to the goals from the nineties [quantum advantage on Shor's factoring algorithm]. I'm following, what are the new goals and how we can make those useful, with the resources we have, with the ideas on the algorithm side that we have.'' 

Educator 9 referred to the simulation of 
interesting phases of matter from QuEra \cite{quera}, and that there will be other algorithms beyond Shor's solving 
useful problems. They said, 
``Whether it's a material problem, whether it's some combinatorial problem...maybe some theorists have a better hunch for what that is, but I would bet that we're gonna have new, interesting problems that quantum hardware will be able to solve that is useful, rather than be able to solve Shor's algorithm in a reasonable time and resources. So, I think by the time we're able to do that [quantum advantage on Shor's algorithm], we would all have abandoned RSA.'' Educator 9 also felt that if we never get an advantage on Shor's algorithm and if the only thing we get out of the [quantum hardware] is ``how we think about information and gain new knowledge and [find that] it has practical implications,'' that this would itself be a great outcome.

\subsection{Q4: Even if you are optimistic about us succeeding at building fault-tolerant scalable quantum computers, do you think we will ever have a quantum computer in our pocket?}

Quantum educators had interesting views about the prospects of quantum computers in our pockets. Some of them thought there was no need for it due to cloud accessibility through a data warehouse, while others interpreted this question literally and discussed the physical constraints of how one would make a quantum computer small and make it work at room temperature to be able to carry it in our pockets. The three codes for this section are 
no need due to cloud accessibility, physical constraints, and running classical and quantum computers in tandem.

\subsubsection{No need due to cloud accessibility}

Some educators noted that quantum computers will be accessed when needed from the cloud on a regular cellphone that one may have in their pocket, similar to the way we access information from the cloud from a data warehouse now. Therefore, having a quantum computer in our pockets will not be needed. For example, educator 2 said, ``It's going to be like a supercomputer, something that sits in a data warehouse and you log into and do your business...”

Educator 6 had similar views, ``We won't need one. In a sense, we won't have a quantum computer in our pocket any more than we have a classical computer in our pocket. If you think about how you use computing power today in your daily life, the near totality of it does not take place in your pocket or on your desk. It takes place in a data center somewhere. But basically, devices you have in your pocket connect to large data centers. What do you do? You browse the Internet, you do searches, you send messages, you write documents, and share them on the cloud. Almost everything you do is not on your physical devices. Your physical devices are just portals to that. So, the future data centers will contain big refrigerators that contain quantum computers, and it doesn't even matter to you that there is a quantum computer. You will notice that certain things become possible [that] weren't possible before. It may be that you still use the same device to connect to them. So, all these ideas, `oh, all these big refrigerators, how am I gonna have one on your desk?' You don't need one on your desk; just like you don't need a data center on your desk. You're already doing it today [with classical computers, where you access data from the cloud].''

\subsubsection{Physical limitations}

Some educators contemplated physical limitations, such as challenges in miniaturization or requirements of cooling to extremely low temperatures, that would make it difficult to have quantum computers in our pockets. Some also questioned why one should worry about having quantum computers in our pocket when we have yet to overcome the challenge of building a fault-tolerant quantum computer.

For example, educator 2 said, “[Classical] computers became so ludicrously powerful compared to everyday needs that they could then add miniaturization. But first, we need it [quantum computer] to exist, and then we're going to miniaturize it. And since we're struggling to get it to exist, or even to define what existence is, the miniaturization is way down my list.”

 Educator 5 focused on challenges for miniaturization and said, ``It could be, maybe, they will never solve the problem [of] how to put it in your pocket...or maybe there's some technology that becomes widely used.'' Their focus remained mainly on the unpredictability of the future, ``Because the other side of the coin is that, if we think about what affects our lives the most today...30 years ago, who could predict that everyone will be on the phones looking at it at all their spare time? So predicting the future, well, it's one of those things where occasionally you get it right, but more times, you get it wrong.''
 
 Educator 7 proposed an approximate timeline for when pocket-sized quantum computers may become possible, while admitting that they were not sure how we would be able to meet all the technological challenges to make it happen. They reflected, “Thomas Watson made the famous statement that the world only needs a [few] computers, and everybody will just connect to it. So let me just put a timeline on that. I think a quantum computer on a cell phone is 50 or 60 years away, and the amount of development that would be needed to get there is unimaginable. We have no idea how that would even conceivably be done. It's kind of like high temperature superconductivity...There might be physics rules against building room temperature quantum computers.''
 
\subsubsection{Running classical and quantum computers in tandem}

Some educators emphasized that classical and quantum computers will always work in tandem, with quantum computers reserved for applications that classical computers can not handle efficiently. Thus, the issue of having a quantum computer in our pockets is of little significance.

For example, stressing that classical and quantum computers will always work together, educator 1 said, ``Just to be clear, everything that comes out is going to be hybrid, because a quantum computer will solve certain things really fast, and in that sense it will always be a co-processor...I don't see this as something that automatically, in the next century or two, takes over from classical computing in the way that we have it.''

Educator 2 echoed similar sentiments, saying, “I think it's asking too much of the [quantum] computer” and “Also, what am I going to use it for?” They continued, “The existence of good classical computers means that you should only use your quantum computer as a last resort. So, do you do so many last resort computing things that you need [a quantum computer] in your pocket? That seems to imply a lot of computational needs of the average human or computational selectivity of the average human [for a quantum computer]. Maybe some limited purposes for cryptography might be a module in our [future] machine, but not a real [quantum] computer [in our pocket], because we're going to use a classical computer for everything we can.''

\subsection{Q5: What is your favorite qubit architecture that is going to play a major role in the future? Why?
}

\begin{figure}
    \centering
    \includegraphics[scale=0.7]{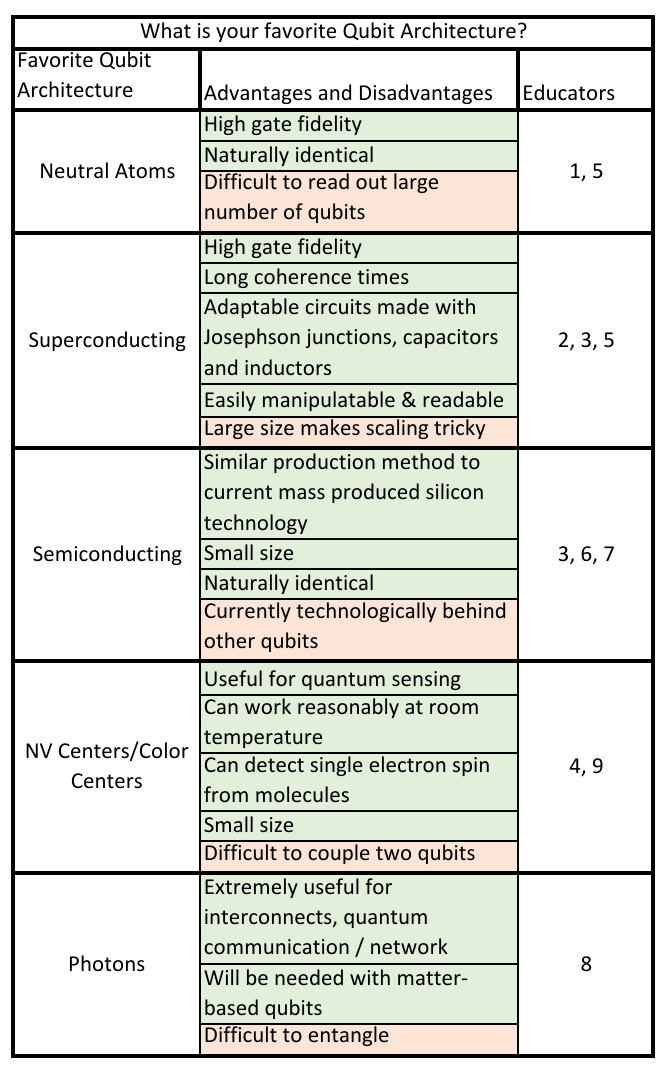}
    \caption{Qubit architectures discussed by educators. Some advantages (highlighted in green) and one disadvantage (highlighted in orange) of each qubit architecture discussed by educators are shown. Educators who selected a given architecture as their favorite are listed in the last column (some educators selected more than one qubit architecture).
    }
    \label{fig:RQ2}
\end{figure}

The coding for this section is based on the different types of favorite qubit architectures selected by the quantum educators. It is important to note that this section only provides an overview of what the quantum educators thought regarding the potential of some of the qubit architectures, and it is not necessarily representative of all of the different qubit architectures currently being pursued. For example, none of the educators specifically selected trapped ion qubits, although they mentioned them particularly while comparing them with their favorite qubits. 
We also note that some of these educators described some features of their favorite qubits 
that were discussed in the preceding sections. 
Fig. \ref{fig:RQ2} describes different favorite qubit architectures discussed, as well as the advantages and disadvantages of each architecture noted by educators.

\subsubsection{Neutral atoms}

Educator 1 reflected on their favorite qubit with future promise, saying, ``On average, over the last few years, my favorite qubit is probably neutral atoms. And let's be clear in expectations and the reason why. It's partly the huge, rapid progress groups in various parts of the world have made...and the gate fidelities are now very high with these things. For many years, neutral atoms lagged well behind other qubit types. The main reason why is because the stability of lasers that you needed to excite atoms with high fidelity to Rydberg states was essentially beyond the technologies that people had in the lab. This has changed rapidly in the last few years. So, you're now getting to two qubit gate fidelities that are starting to become competitive with superconducting qubits. And neutral atoms have the same advantage that trapped ions have over the solid-state qubit types; that they don't rely on the material science and the fabrication, and you have naturally identical qubits simply by taking atoms and using atoms that are initially identical. It gets rid of a lot of calibration problems and a lot of the difficulties in scaling up in terms of the sheer number of qubits that you can produce, track, and work with.''

Educator 1 continued, ``The advantage that neutral atoms have over trapped ions is that when you add an extra qubit to your system, it doesn't change the way in which you do your gates. In trapped ions, people need to find all these ways to do photonic interconnects or shuttling of ions around the chip; because when you do interactions, you have to have a small number of ions, if you want to do high fidelity two qubit gates, because you are using the motional state, motional chain as a bus for the quantum information basically. The difficulty with this is just that if you add an extra ion, it changes all of the collective motion and excitations. So, you don't have that problem with neutral atoms. Instead, you can take these things, and you can scale up relatively straightforwardly to at least of the order of a few 1000 physical qubits. And so, if you ask me which system is going to most easily get to a couple of 1000 qubits, I would say that neutral atoms are highly competitive against the other two leads at the moment, which are really trapped ions and superconducting qubits, the way I see things.'' 

Educator 1 then reflected on other qubit architectures and noted, ``There are other really interesting quantum systems that have been developed in different parts of the world. There are certain things that have been done with photons that are...quite interesting. We have difficulty as to how you essentially do measurement and feed forward [use the results of measurements to perform subsequent gate operations]. And silicon qubits, of course, are interesting. But apart from the level of control you need to gain to really start scaling up [to] a massive scale, they're, of course, exciting [since they are] so close to the mass-produced silicon technologies that we use today...But if you ask me for my favorite at the moment, it's neutral atoms. But for a theorist [like me], there is really exciting stuff being done across so many different platforms that it's a fun time.''

Regarding their favorite qubits, educator 5 noted, “more traditional stuff like cold atoms, superconducting qubits….they seem much more promising right now.”

\subsubsection{Superconducting qubits}

Regarding their favorite qubit that they think has promise for the future, educator 2, who works on superconducting qubits, mentioned how superconducting qubits they use today are different from a decade ago, as they have figured out  
how to fix certain problems. They explained, 
``For instance, charge noise is really terrible. So, we invent qubits that are not sensitive to charge noise. So, for superconducting circuits, since everything is very synthetic, and we just build it, yes, the qubits may change. They'll definitely adapt, as we learn more about one of the problems that are hard to solve, we will adapt the qubits to be less sensitive to the things that we really can't handle. So, for us, it's not that big of a deal if you go from fluxonium to transmon, or transmon to [another] qubit. If you look at it from a distance, it has some inductors, capacitors, and Josephson junctions, and it just stays that way. Even fancy protected qubits are just the same materials that we've assembled in a clever way.'' Educator 2 explained that as qubits continue evolving, since they are all made from the same underlying material, it is easily possible to replace an older type of qubit with a newer one.
They continued, ``It's harder for me to judge outside of my field. Certainly, neutral atoms are coming really far, really fast.''

Educator 3, who has done research related to semiconducting qubits, identified them and superconducting qubits as their favorite qubit architectures. Regarding why they prefer superconducting qubits, they said, "Superconducting qubits are very impressive in terms of their coherence times, and ability to manipulate and read out. But they are big so that kind of cuts into their scalability. But that doesn't mean that it's not possible [to build scalable quantum computers using superconducting qubit platform]." 

Educator 5 expanded on why they like superconducting qubits, saying, “I don't have a very good sense of all the different platforms, because I basically only collaborate on one of them, which is superconducting qubit. So that's kind of the area that I probably know a bit more about, what has been done [in it] and where it's going in comparison to other platforms. So my knowledge is very much informed by what I've been exposed to and what I've done work on…I'm obviously most excited about the stuff I'm working on which is...really superconducting qubits.”

\subsubsection{Semiconducting qubits}

Educator 3 said that they like both semiconducting and superconducting qubits.  
About why they like semiconducting qubits, they said, ``that's the first one I thought about [for my research]."

Educator 6, who works on semiconducting qubits, reflected on why they find them promising. They said, ``So in terms of quantum computing...the spins of the atoms would be quantum bits, and we are looking at connecting them together, making entangling quantum logic operations between them by using the interaction between the electrons. Because all of this is built inside of a silicon chip, which is fabricated with the same manufacturing principles that are used to fabricate the chips that are in your computer and mobile phone, the idea is that if we can get the basic physics right, then the manufacturing can be done using the trillion-dollar semiconductor industry that already exists. We don't need to reinvent the wheel. Of course, we need to adapt the wheel because there are important details that need to change. But fundamentally, the platform is Silicon CMOS, which is the most successful engineering feat that humanity has ever created, right?''

Educator 6 further contemplated on semiconducting qubits, ``There are now companies entering this field, and one of them is Intel, famous for its semiconductor chips. They started some 5--6 years ago a semiconductor qubit program, and they now have significant results in this field. And now through an initiative of the U.S. Government, something called the Qubit Collaboratory, they are now sending out these semiconductor qubit devices that they make in their semiconductor foundry to academic research groups. This was something that the U.S. funding agencies recognized early on---that the barrier to entry in semiconductor quantum technologies was so high for academic groups that it was actually stifling creativity. Innovation was too hard for academics to make those things. So, they've put together this consortium, where, for example, Intel makes these structures and then dishes them out to academics to do quantum science with them...''

They continued, ``So, no existing quantum computer is useful....They need to grow by orders of magnitude in terms of number of qubits, and in terms of quality of qubits before they become useful. So the hope here is that the extraordinary process control and existing infrastructure that underpins silicon electronic technology can come to the rescue for quantum technology. The problem has always been how to make that transition from academia to industry. So there had to be some pioneers, sort of cowboy academics, who actually tried to make those silicon qubits. And we, for a decade, for 15 years, we were the stray dogs of the field. We saw people [who work on] superconducting qubits making all fantastic stuff, and we were still tinkering in the clean room with devices that didn't work. Now we got them to work. We are at 99\% fidelity operations on everything. So, we are now
where superconductors were 10 years ago. But in the long term, the vision is that once we get the trillion-dollar semiconductor industry to underpin the growth from here, then we can take over.''

Educator 6 added, ``By the way, we're not the only ones thinking this way. There's a company, you may have heard of them, called PsiQuantum \cite{psiquantum}. It's a very successful company in Silicon Valley that makes photonic quantum computers. They understood straight away, if we want to make a useful quantum computer, we need to use silicon CMOS technology. So, they make their photonic chips in tier one semiconductor foundries. These things are extremely expensive. We're talking hundreds of millions of dollars of cost just to get those chips off the ground using those facilities. But that's how you get the scale and the reproducibility.''

Educator 6 further acknowledged that the other advantage of semiconducting qubits is the ``size...that's the other thing, semiconductor qubit, it's small, the qubits that we make are actually atomic size. The structures around them are the size of a transistor. But you know how many transistors we have in a chip nowadays, about 10 billion on a chip this size [Educator 6 gestured with their hand, indicating that the chip is a few centimeters squared]...[so] spins in semiconductor devices are very promising. That said, superconducting qubits are great as well, and ion traps are making really great progress in shuttling things around. So, I don't think there is any sense in which anyone can point the finger at all, [saying] `oh this is by far the favorite contestant'.  
But when you look at the integrated trap ion traps, they are basically silicon CMOS devices themselves. So, the division is quite blurry. 

Educator 6 continued, ``Some of the superconducting qubits are built within facilities and with methods that are not very different from semiconductor qubits. So, people are getting very creative at getting the best out of what exists and then inventing new things when necessary. My own favorite system is actually the antimony atom, not phosphorus, because antimony has a spin 7/2 nucleus, which is an eight-level system. So, it actually encodes three qubits in the nucleus. Antimony atom is actually a four-qubit system, three in the nucleus and one in the electron. So, it's very high density of information, and it allows all sorts of interesting things. You can do some embryonic quantum error correction within the nucleus itself, just in one nucleus. So that's the thing I'm actually investing on right now...basically, you make a mini logical qubit in the nucleus itself without having to couple to anything, so the nucleus already has enough dimension of Hilbert space to afford some of those quantum error correction operations.''

Educator 7 also noted that semiconducting qubits were their favorite (despite working in quantum optics), saying, ``The beauty of the system is that you have silicon technology, which is always used in ordinary computers, and the technology, the materials engineering, is very well developed and they can make highly pure silicon crystals. They know how to pattern them and make circuits and all the microstructures you need.'' Going into details about a particular type of semiconducting qubits in which they see great potential they said, ``They take a single phosphorus and put them exactly where they want it, and then they embed another single phosphorus atom, and then another one, another one exactly where they want them. So, the qubits are single atoms, but in silicon. This is, in a way, the best of all worlds, because you have single atoms which are perfectly identical, so that's like trapped ions. But then you're dealing with atomic systems, which are more in a way pure than say, quantum dots or man-made, human-engineered microstructure, which are never perfect. It might take 20 years for them to actually do all the technical steps and learn all the physics they need to do error correction on these phosphorus qubits.''

\subsubsection{Nitrogen-vacancy centers/Color centers}

Educator 4, who works on nitrogen-vacancy (NV) centers in diamonds, saw great promise in that qubit architecture \cite{gurudev,nvcenter1,nvcenter2,nvcenter3}, particularly for quantum sensing. They said, ``The nice thing about them is that because they are sitting in a diamond crystal, you can think of them as being somewhat isolated from the rest of the world, because diamond is one of the hardest materials. But it's also got many other amazing properties for physics as well.'' They noted that NV centers are defects, i.e., alterations of the local environment, and they can have some nice applications due to their quantum properties saying,
``The application of these properties...is one to study things like very small volumes of samples, like a two-dimensional material on the surface, or to study molecules that we place on the surface of the diamond. And then you can imagine our qubit, which can be less than a few nanometers away, a billionth of a meter away from the surface, can be used to then do spin resonance measurements on the molecules at very small volume...we can detect down to a single electron spin from those molecules, for example. And this kind of...sensitivity is a really big step forward...as we make our system more robust, more stable, able to take large amounts of data, we start to incorporate things like, biologically relevant molecules. And we want to try and study them at room temperature, wet conditions, something that really no technique can actually do until now.''

Educator 4 further noted that the NV center in diamond is a valuable qubit that can be useful for probing foundations of quantum physics, e.g., quantum macroscopic superpositions. They said, ``...I'm sure you have heard of the Schr\"{o}dinger cat experiment. The thought experiment---the idea of the Schr\"{o}dinger cat was, we take a macroscopic object and why don't we see it existing in a superposition of, say, dead and alive at the same time. Experimentally, most macroscopic objects, they tend to be very clearly in one position or one momentum state, or one energy state, or whatever it is that you're probing. They tend to be in that state and not existing in superpositions. There have been some theories for a long time about why that might be, including things like gravity that decoheres quantum superposition. So, imagine you have a superposition of a particle that's in, say, position A and position B, then gravity somehow breaks that superposition once the superposition gets large enough. So that's one theory, for example, that's been out there which is proposed by Penrose many, many years ago. Similarly, there are other theories which say, well, actually, it's some kind of nonlinearity in the Schr\"{o}dinger equation that only shows up at those length scales...So, there are many people with many theories, although these are 2 probably primary theories that people are interested in. And so, [NV centers provide] one way to prove these theories, and to see if there are indeed modifications of quantum mechanics at this intermediate landscape. So, I'm not talking atoms, molecules, we understand quantum mechanics really applies very well there [at those length scales].''

Pointing to the challenges laid out in the DiVincenzo criteria~\cite{divincenzo}, educator 4 noted, ``We want very coherent things because we want them to last a long time, so that we can do all the quantum manipulations on them. But at the same time, if you make them too isolated, then it's hard to create interactions between them. So, one of the focuses that people obviously have to figure out is, how do you engineer the mechanical structure in such a way that you cut out all the bad interactions but only leave behind the good interaction between the two oscillators, such that maybe they are at two resonant modes.''

Educator 9, who works broadly on color centers, pointed to color centers as their preferred qubit saying, ``Superconducting qubits, they're ahead of everyone. They've been [important] in terms of making the NISQ era happen'' but ``what I love about color centers is, whenever I make a trip, I can just bring it to someone, and they can put it in their cryostat and measure it and use it...That's not the case with trapped ions, even though they're also pretty cool qubits...''

\subsubsection{Photonic qubits}

Educator 8, who does theoretical work on photonic qubits, noted, ``I know that some of these platforms I work on in optics, it is challenging to build, to prepare these multi-qubit photonic entangled states. Theoretically, protocols have been schematically worked out actually, but to implement...these multi-qubit entangled states with single photons is currently challenging. There are companies that are working on it. So, what I'm trying to get at is that some of these scalability issues are probably far easier to tackle with photons, but they come with their own challenges. You can prepare photons in millions, but then to entangle them...And people are trying to figure out efficient ways of building large, entangled states...That's why I personally feel as many people do, there is no front runner [qubit] yet. Matter-based systems [as qubits] are nice to study, but scaling them up is going to be pretty hard.''

\subsubsection{No clear winner}

Several educators emphasized that, despite expected advances in quantum computing over the next few decades, different qubit architectures would likely continue to have certain advantages so that we will continue to harness them for various applications tailored for the affordances they offer. Thus, no single qubit architecture could be regarded as a definite winner, making it essential to conduct research on all qubit architectures at this early stage of development. The educators also stressed that advances in quantum computing, sensing, and networking will be synergistic and mutually beneficial.

For example, educator 1 emphasized that there are different applications that one should keep in mind, saying, ``so I think we need to be aware of the variety of different applications here. And I think there may be an architecture for fault tolerant digital quantum computing that emerges as the strongest but I think that irrespective of which one emerges in that form, photons will always be there for interconnects and for things like quantum communications and networking. And so, I think that will always have its importance.''

Educator 1 expressed optimism that analog platforms will continue to excel at certain kinds of quantum simulation, saying, ``I think there's always going to be an opening for analog platforms for doing quantum simulation, because I think for many, many decades to come, we will have analog devices that can do quantum simulation better than you can on a digital machine.'' They made an analogy with engineering practices, and that even though there are now highly advanced digital codes for  
hydrodynamics and aerodynamics, wind tunnels and water tunnels are still being used, as building scale models and sticking them, e.g., in a wind tunnel is still better for certain purposes.
They continued, ``I think we can see the same on the quantum side, that there are certain types of quantum simulation that will always be better on an analog platform, and that will then depend very much on the problem being solved, and what 
you can access most easily in a particular type of hardware.''
Regarding the strengths of different quantum platforms for measurement and sensing applications, educator 1 said,  ``When it comes to measurement and sensing, again, I still think you're not going to go past atoms or ions for atomic clocks. Also, we bring in things like nitrogen-vacancy centers and spin qubits...different forms to do close up, to do measurement and sensing. So, I think all of these platforms across a range of quantum technologies are going to have their advantages and uses.''

Educator 1 also stressed that it is important that a variety of architectures and applications are simultaneously being pursued, ``There are challenges that all of these different technologies have, that they have to overcome. And there can be a few changes in design that suddenly make a huge difference and lead to a massive advantage for one, or maybe for multiple platforms. And I think we are just gonna have to wait and see what happens. It's exciting to see people working on such a range of different things...Any of these things give you the entangled quantum systems that could really be used in different contexts. And I think that you'll see a lot of that happening, things that are being developed for quantum computing being translated into quantum communications, sensing, metrology and so on in a way that again can bring a lot of benefits. And so, I think that I don't feel that, at this stage, we need to pick a winner in the platforms. I think that actually, there is a lot more to be gained in the technological development in this field by continuing with all of the platforms and looking at where the science and engineering take us.''

Educator 8 stressed that they felt that there would not be one winner, saying, ``I really think we are not going to have one winner...photonics has to be a winner for us to network, anyway...So photonic qubits will exist in the future quantum computer for them to do quantum network, quantum Internet. But from matter-based qubit point of view, I think, every system seems to have its own advantages and disadvantages...we know even right now that we can most likely envision a hybrid quantum network in the future, like specific types of qubits for specific nodes. So, I think all the research is valuable right now, maybe some of them [qubits] give way to better materials, like people may discover better...solid state qubits...But I think both matter-based and photonic qubits are here to stay.''

\section{Summary and Discussion}

Our findings reveal that among leading quantum researchers who are educators we interviewed, there is consensus regarding many fundamental questions about quantum technologies that are often asked by students and others, although there appear to be some differences in the timeline for when a particular goal (e.g., quantum advantage on Shor's algorithm) will be achieved in the future, as well as their favorite qubits with future potential. 

\subsection{Quantum computing's current state}

On the question of whether we currently have a quantum computer, there is broad agreement that we do have quantum computing devices in the NISQ era, though their capabilities remain very limited compared to the ultimate goals envisioned in the future. The analogy to early classical computers, offered by educator 1, resonates throughout the responses. Quantum educators agreed that the current quantum computers can be viewed as the vacuum tube stage of a technology that will continue to evolve dramatically.

Quantum educators acknowledged that current NISQ-era devices successfully create superposition, entanglement, and interference for performing calculations. However, the educators' perspectives highlight an important distinction between having quantum computers that demonstrate quantum mechanical principles and having practically useful quantum computers. They emphasized that the current devices fall far short of the fault-tolerant, scalable systems.

\subsection{Timeline uncertainties and the path to small and scalable fault tolerant quantum computers}

Another important finding is that 
most quantum educators were optimistic that we will have small fault-tolerant quantum computers in a time-scale of around 
10 years. The timeline for a scalable fault-tolerant quantum computer, which can execute transformative applications like breaking RSA encryption using Shor's algorithm with quantum advantage, varied (although most noted it would not be achieved in the next few decades). Educator 6 noted it could be ``the most exciting thing'' if we discover a fundamental law preventing such a quantum computer. This variation reflects the genuine uncertainty in the field about overcoming major technical challenges and achieving these long-term goals.

Several educators emphasized that progress in QIST may not be linear. For example, educator 2's observation about finding ways to reduce the overhead for making logical qubits suggests that breakthrough innovations could accelerate timelines. Conversely, educator 5's comparison to the slope in classical computing's Moore's law \cite{moore} (e.g., depicting the timeline for how quickly progress is made) suggests that quantum computing progress may follow a ``much shallower slope.'' This prediction suggests that patience is required to achieve major milestones in quantum technologies.

\subsection{Applications beyond Shor's algorithm}

An important theme throughout the interviews is the emphasis on applications beyond Shor's algorithm for factoring products of large prime numbers, which led to investments in QIST due to its implications 
for encryption and national security. Multiple quantum educators (e.g., educators 1, 4, and 9) highlighted quantum simulation as likely to provide practical value, e.g., before code-breaking and other classical applications with quantum advantage. For example, many quantum educators focused on materials science, drug discovery, and optimization problems 
as domains where quantum computers might first demonstrate practical advantage.

\subsection{Quantum computers as infrastructure, not personal devices in pockets}

The agreement that quantum computers will not become personal pocket devices represents a significant departure from how the trajectory of classical computers has unfolded. Some educators' responses reflect a concern for both the challenges involving physical requirements of quantum computers (e.g., how they can be miniaturized and made to work at room temperatures)
and the computational landscape of the 21st century. Educator 6's observation that we already do not have classical computers in our pockets, just portals to cloud computing, provides a meaningful approach for understanding quantum computing's future in this regard. Quantum educators also emphasized that quantum computers will only be used for applications for which they are significantly better suited than classical computers due to their quantum advantage.

\subsection{Architectural diversity and no clear winner}

The responses to the question about favorite qubit architectures that show promise reveal a diversity of approaches in the field. While most researchers naturally showed a preference for their area of expertise,  
educators generally believed that no single qubit platform has emerged as the clear winner. Neutral atoms (educators 1 and 5), superconducting circuits (educators 2, 3, and 5), semiconducting qubits (educators 3, 6, and 7), color centers/NV centers (educators 4 and 9), and photonic systems (educator 8) all have advocates who articulated their specific advantages.

Furthermore, this diversity in qubit architecture is viewed positively by the educators, with many emphasizing that different platforms may ultimately serve different purposes (or even be combined in some ways in the future). For example, educator 1's point that ``there is a lot more to be gained in the quantum technology development in this field by continuing with all of the platforms'' reflects their deep understanding that the quantum technology ecosystem will likely be heterogeneous (at least for several decades to come), with different architectures potentially optimized for different applications.

\section{Implications for Education, Outreach, and Communication}

These findings have important implications for how quantum educators should communicate with students about quantum computing as well as for curriculum development. In most physics education research that focuses on specific content areas, the physics itself is well established and uncontroversial, e.g., in classical mechanics or standard quantum mechanics. In that context, the unresolved questions in physics education research instead center on how students learn, interpret, and make sense of concepts.
In quantum computing, the content is rapidly advancing, so taking a snapshot of expert consensus, or lack thereof, is important as a baseline for other education research along these strands and curriculum development.

Since preparing students for the second quantum revolution is critical, it will be valuable to incorporate the issues discussed here not only in interdisciplinary QIST courses but also in modern physics and quantum mechanics courses for physics majors, ensuring that students do not develop a distorted view of the current status and future prospects of quantum computing. Moreover, not all educators who teach modern physics and quantum mechanics courses are QIST researchers, so these findings are particularly valuable for guiding how they may discuss these issues with their students.

One framework educators can adopt to discuss these issues is the comparison of classical and corresponding quantum concepts \cite{singh2022tpt}. For example, in classical computing, error correction may involve making copies of bits. One can use three classical bits instead of one, and if one bit flips due to error, the majority value (either 0 or 1) is taken as the correct value. Quantum educators can then discuss that the state of a qubit cannot be cloned due to the no-cloning theorem \cite{nocloning}. Consequently, quantum error correction is, in general, very difficult, and therefore making a fault-tolerant quantum computer is extremely challenging. Although teaching complex quantum error correction codes \cite{shorcode,surfacecode} may not be appropriate in many courses, for example, in a quantum mechanics course for physics juniors and seniors, the no-cloning theorem is relatively accessible, and students can learn it to gain insight into the challenges of quantum error correction and building fault-tolerant quantum computers. 

Education researchers developing modules for interdisciplinary QIST courses 
should consider where insights provided in this paper related to students' common questions would fit best and embed these findings in the curricula developed. This is relevant for courses for students across different science and engineering majors, as well as for physics majors in modern physics or quantum mechanics courses.

Historical analogies and comparisons with early classical computers provide accessible paradigms for helping students learn about current limitations and future potential of quantum technologies. For example, using analogies relating to vacuum tube technology, which was used for classical computing in early stages, can be a helpful reference for students to understand the current limitations of quantum technologies. Research in the quantum education context suggests that using elaborated analogies can be highly effective because they allow students to map known concepts into more recent or abstract ones \cite{faletivc2025analogies, didics2015analysis}. References relating to classical computers that previously required extremely large amounts of physical space in order to complete calculations, which can now be completed with a modern smartphone, can provide useful context when addressing students' question about the future potential of quantum technologies, including how they are likely to evolve.

Educators should also emphasize near-term applications of these quantum technologies. While discussions of future potential applications like code-breaking and other classical applications with quantum advantage can be exciting, they can also reinforce unrealistic expectations for the short term. This aligns with recent studies on hype regarding quantum technologies, suggesting that students often arrive in the classrooms with certain conceptions based on popular media accounts \cite{meyer2023media}. Also, since quantum simulation will show benefits before circuit-based quantum computers, educators may, e.g., find it helpful to instead discuss applications relating to quantum simulation and optimization in order to set proper expectations for students in the near term. This can be further emphasized by noting how fragile qubits are, and attempting to scale quantum technologies can introduce unexpected difficulties, as quantum educators in this study pointed out. In particular, improving on a particular aspect of a quantum system without making other aspects worse can be extremely challenging due to the fragility of qubits. This type of discussion may provide good context for students to understand the difference between what are near-term application goals and what are further in the future.

It may also be helpful to discuss with students that the term ``quantum computer" is now being used quite flexibly, and any machine that does computation and uses quantum mechanics principles can be called a quantum computer.
Educators may discuss with students that working on NISQ-era non-fault-tolerant quantum computers can be useful for learning about quantum systems, even if it is primarily for research purposes, e.g., as pointed out by educator 7 in the context of car companies conducting various tests using current NISQ-era quantum computers.

Furthermore, the cloud-computing model for quantum access can help students and the public understand how accessibility for future quantum technologies may look. This can assist with understanding that future quantum technologies may not look like a quantum computer in your pocket, instead, it may be something which you connect to on the cloud. Cloud-based models have been highlighted as a useful pedagogical tool that allows students to access remote quantum experiments \cite{borish2025affordances}. It may be helpful to make students familiar with cloud access and to engage them with quantum experiments remotely in these early stages.

The quantum educators' views on timelines should be transparently communicated to help students, the public, and the media understand that uncertainty is inherent in predicting technological development. This would assist with the previous points relating to providing realistic expectations. 

It is also important for these uncertainties to be communicated alongside clear explanations about current capabilities and limitations. Doing so not only provides valuable context to students who are being introduced to quantum technologies and may have questions about them, but also helps build public trust and understanding. 

Furthermore, it will be useful to have discussions with students about the fact that researchers think that, despite the potentially long timeline for building a scalable fault-tolerant quantum computer, the work is worthwhile to continue exploring. Many researchers noted that there is much to learn about new applications in which quantum technologies can be utilized. These uses are not necessarily constrained to quantum computing, and these technological advancements can potentially have spin-offs that can help to transform current technology. Therefore, a good context to provide to students is that any work they do in this field can be helpful in many ways beyond just building a quantum computer.

\section{Conclusions}

Although the number of quantum educators was limited, this study captured perspectives from leading quantum researchers, who are educators on fundamental questions about quantum information science and technology that students and the public often ask. These expert reflections provide
valuable insights for educators, science communicators, and policymakers seeking to convey accurate information about quantum computing's current state and future prospects. These findings can be valuable for developing curricula to help students at different levels learn these concepts.
Below, we summarize our key findings that can help educators in both formal and informal educational settings, in addition to helping the policy-makers and other stakeholders.

Regarding current machines in the NISQ era, consensus exists that we do have quantum computers, though they are analogous to early classical computers, i.e., functional but limited in capability. Current NISQ-era devices demonstrate quantum principles but lack the fault tolerance and scale needed for transformative applications.
Regarding timeline uncertainty, expert estimates for achieving small fault-tolerant quantum computers show more certainty than their estimates for scalable quantum computers that will show computational advantage on Shor's algorithm. This  
reflects genuine uncertainty about overcoming technical challenges in developing scalable quantum computers, even if we are able to build small fault-tolerant quantum computers within a decade. This 
finding related to the timeline uncertainty should be transparently communicated to students and other stakeholders to avoid creating unrealistic expectations.
Regarding whether we will have quantum computers in our pockets, experts pointed to a cloud-based future similar to classical computers, emphasizing that quantum computers will remain specialized tools accessed through cloud infrastructure rather than personal devices. This reflects physical constraints as well as the modern computing paradigm of centralized processing power we already have for classical computers.

Experts noted that near-term applications of quantum computers will likely be in quantum simulation, e.g., for applications in materials science, drug discovery, and optimization problems, which will likely provide practical value before applications like quantum advantage on Shor's famous factoring algorithm, offering more realistic near-term goals for quantum technologies.

Regarding quantum educators' favorite qubits with future potential, no single qubit platform emerged as the clear winner,
especially because we are at the initial stages in the development of quantum technologies. Quantum educators noted that neutral atoms,  
superconducting qubits, semiconducting qubits, photonics systems, and other approaches each show promise, and the future quantum ecosystem, at least in the next few decades, will likely be heterogeneous with different architectures serving different purposes. Moreover, photonics will always have a role in quantum networks.

These findings suggest that at this early time in the development of quantum technologies, quantum computing and quantum information education should emphasize historical context, acknowledge uncertainty, celebrate diversity in approaches with many qubit architectures being pursued simultaneously, and set realistic expectations about applications and access models. By grounding student and public understanding in expert perspectives, we can foster informed enthusiasm for quantum technologies while avoiding the pitfalls of hype and misinformation.

As QIST continues to evolve, periodic reassessment of expert perspectives will be valuable to clarify student and public doubts. It can provide guidance to other educators in addressing these types of questions from their students and developing curricula to address these issues. The field's rapid progress, combined with fundamental uncertainties about ultimate capabilities and the confusion surrounding these issues, requires ongoing dialogue of the type captured in this research.
This study provides a snapshot of expert thinking at a critical juncture in the development of QIST, offering guidance to educators and other stakeholders for understanding and navigating the complex landscape of quantum technologies.

\section*{Acknowledgments}
We are grateful to all educators who participated in this research. This research is supported by the U.S. National Science Foundation Award No. PHY-2309260.


\section{References}

\bibliography{refs}

\end{document}